\documentclass[showkeys,showpacs]{revtex4}
\usepackage{amsmath}
\usepackage{amssymb}
\usepackage{graphicx}

\begin{document}

\title{Some thick brane solutions in $f(R)$-gravity}

\author{
Vladimir Dzhunushaliev,$^{1,2,3}$
\footnote{
Email: vdzhunus@krsu.edu.kg}
Vladimir Folomeev,$^{2,3}$
\footnote{Email: vfolomeev@mail.ru}
Burkhard Kleihaus,$^{3}$
\footnote{Email: kleihaus@theorie.physik.uni-oldenburg.de }
Jutta Kunz$^3$
\footnote{Email:  kunz@theorie.physik.uni-oldenburg.de}
}
\affiliation{$^1$Department of Physics and Microelectronic
Engineering, Kyrgyz-Russian Slavic University, Bishkek, Kievskaya Str.
44, 720021, Kyrgyz Republic \\ 
$^2$Institute of Physicotechnical Problems and Material Science of the NAS
of the
Kyrgyz Republic, 265 a, Chui Street, Bishkek, 720071,  Kyrgyz Republic \\
$^3$Institut f\"ur Physik, Universit\"at Oldenburg, Postfach 2503
D-26111 Oldenburg, Germany
}

\begin{abstract}
The thick brane model is considered in $f(R)\sim R^n$ gravity.
It is shown that regular asymptotically  anti-de Sitter solutions
exist in some range of values of the parameter $n$.
A peculiar feature of this model is the existence of a fixed point
in the phase plane where all solutions start,
and the brane can be placed at this point.
The presence of the fixed point allows to avoid fine tuning
of the model parameters to obtain thick brane solutions.
\end{abstract}

\keywords{Thick branes; Higher-order gravity theories}

\maketitle

\section{Introduction}
At the present time, the study of the structure and evolution
of the Universe is at an interesting stage.
The necessity of a consistent description
of the present Universe demands the creation
of a unified theory of elementary particles and cosmology.
A very promising way is the consideration
of models of the Universe in higher-dimensional theories.
Such investigations were initiated in the works of Kaluza and Klein
in the 1920s for the unification of the two fundamental interactions
known at that time
- gravitation and electromagnetism -
within the framework of a unified five-dimensional theory.
Later on, similar ideas were used for the unified description
of the four currently known fundamental interactions
within the framework of superstring theories
with several extra space dimensions.
As in the case of the Kaluza-Klein theories, in superstring theories,
it is supposed that our four-dimensional space-time results
from the spontaneous compactification of a higher-dimensional space.

At the same time, models of the Universe with non-compact
(and even infinite) extra dimensions
are under consideration \cite{randall}
(for a review, see also \cite{Rubakov:2001kp,Barvinsky:2005ak}).
In such a theory, it is supposed that
we live on a thin leaf (brane)
embedded into some higher-dimensional space (bulk),
and matter is somehow confined (trapped) on the brane.
The existence of extra dimensions then allows to resolve
a number of old problems in high-energy physics
(such as the problem of mass hierarchy, stability of the proton, etc.).

All branes can be divided into two classes: thin branes and thick branes.
In the first case, one has a delta-like localization of matter
on the brane \cite{randall}.
From a realistic point of view, however,
the brane should have some thickness.
The inclusion of a brane thickness then
yields new possibilities and new problems
(for a review, see, e.g., \cite{Dzhunushaliev:2009va}).
Such a brane must satisfy two major requirements:
1) the solutions should be regular and asymptotically flat,
or de Sitter ones (or anti- de Sitter ones);
2) ordinary matter should be confined to the brane.

Most thick brane models employ scalar fields
within the framework of Einstein's theory of gravity
(see e.g.~the review \cite{Dzhunushaliev:2009va} and references therein).
However, one might expect the existence of brane-like solutions
also for some kinds of modificated gravity theories,
the so-called higher-order gravity theories (HOGT).
In such theories the action of
the Einstein-Hilbert gravitational Lagrangian is supplemented
by further terms,
which are curvature invariants~\cite{Sakharov:1967pk}.
(Such a modification is based on the effect of the interaction
of quantum matter fields with the classical gravitational field.)
This allowed to avoid an initial cosmological singularity
and to construct regular cosmological models of the early Universe \cite{Ruzm}.
Later it was shown that in such type of models a stage of inflation
can exist \cite{Starobinsky:1980te}.

Currently, this last possibility is widely used
for the description of the present accelerated expansion of the Universe.
This acceleration can be explained by the presence
of some antigravitating substance - the so-called dark energy.
The description of dark energy can also be realized within
the framework of $f(R)$ theory, where $f(R)$
is some arbitrary function of the scalar curvature $R$.
By choosing $f(R)\sim R^n$, it was shown that such models are in good
agreement with several different sets of observations~\cite{Capoz}-\cite{Noj}.
On the other hand, such theories can be successfully employed
for the description of dark matter~\cite{Capoz1} as well.

HOGT with more complicated combinations of curvature invariants
are also under consideration.
In particular, in the low-energy limit of M-theory
the Gauss-Bonnet invariant appears
$$
G=R^2-4 R_{\mu\nu}R^{\mu\nu}+R_{\mu\nu\rho\sigma}R^{\mu\nu\rho\sigma}.
$$
It was shown that such models, on the one hand,
do not contradict observations within the Solar System and, on the other hand,
successfully describe the present accelerated expansion
of the Universe~\cite{Noj1}.
These models can be used in the description of
an effective equation of state
both for an effective cosmological constant
and for the dynamic case (quintessence, phantom dark energy),
and also for the description of the transition of one type of dark energy
(quintessence) into another one (phantom energy).
Also there are theories which employ both $f(R)$
and Gauss-Bonnet terms to describe dark energy~\cite{Noj2}.

Another usage of HOGT consists in the consideration
of higher-dimensional cosmological and astrophysical models.
In particular, in Ref.~\cite{Noj3} brane world
and black hole models with higher-dimensional action
$$
S=\int d^d x \sqrt{-g} \left[aR^2+b R_{\mu\nu}R^{\mu\nu}+c R_{\mu\nu\rho\sigma}R^{\mu\nu\rho\sigma}+\frac{1}{k^2}R-\Lambda+L_m\right],
$$
(where $L_m$ is the matter Lagrangian; $a,b,c$ are arbitrary constants)
were considered.
The obtained results allow to estimate
general properties of models within the framework of HOGT.

Furthermore there are some results obtained by the application of HOGT
for the creation of brane-world models \cite{Noj4}.
In particular, Parry et al.~\cite{Parry:2005eb}
considered a brane model in $f(R)=R+\alpha R^2$ theory.
Making use of the conformal equivalence of such gravity models
and Einstein-Hilbert gravity with a scalar field,
the authors rewrote the $f(R)$-equations in the form of the
Einstein equations with some scalar field source.
They showed that in such models brane-like solutions exist.

Whereas Parry et al.~\cite{Parry:2005eb} considered thin branes,
in this paper we investigate thick branes in such
$f(R)$ theories to see, whether this leads to new
and physically more acceptable results.

\section{Equations and solutions in $f(R)\sim R^n$ theory}
We will work in a five-dimensional spacetime. The corresponding
gravitational action can be taken in the form
\begin{equation}
\label{5daction_f}
S = \int d^5 x\sqrt {-^5g} \left[ -\frac{R}{2}+f(R)
\right]~,
\end{equation}
where $f(R)$ is an arbitrary function of the scalar curvature $R$.
(Here we employ units such that $8\pi G=c=1$.)
Variation of the action \eqref{5daction_f} with respect to the
$5$-dimensional metric tensor $g_{AB}$ led to gravitational
equations:
\begin{equation}
\label{einst}
R_{A}^B-\frac{1}{2}\delta_{A}^B R=\hat{T}_{A}^B,
\end{equation}
where capital Latin indices run over $A, B,... =
0, 1, 2, 3, 5 $, and
\begin{equation}
\label{T}
\hat{T}_{A}^B=-\left\lbrace \left( \frac{\partial f}{\partial R}\right) R_{A}^B -
\frac{1}{2}\delta_{A}^{B} f+\left( \delta_{A}^{B} g^{LM}-\delta_{A}^{L} g^{BM}\right)
\left( \frac{\partial f}{\partial R}\right)_{;L;M}\right\rbrace
\end{equation}
specifies the effective geometric matter source with the nontrivial
dependence on curvature; the semicolon denotes the covariant derivative.
One can see that the gravitational equations in
$f(R)$-gravity rewritten in the form \eqref{einst}
have a structure that coincides with the standard general relativity equations
when the source of the gravitational field is
the effective energy-momentum tensor \eqref{T}.
One can check that the energy-momentum conservation law
is satisfied as well (see, e.g., \cite{Ruzm}).

Here we will focus on a special choice of $f(R)$ in the form
\begin{equation}
\label{fR}
f(R)=-\alpha R^n,
\end{equation}
where $\alpha>0$ and $n$ are constants.
As shown in Refs.~\cite{Capoz}, where the present
accelerated expansion of the Universe was considered,
there are some ranges of $n$
which do not contradict the observational cosmological data.
Therefore it seems natural
to consider these values of $n$ for brane models as well.

We will adopt the flat brane model with the metric
\begin{equation}
\label{metric}
ds^2=e^{2 y(z)}\eta_{\alpha \beta} dx^\alpha dx^\beta-dz^2,
\end{equation}
where the warp factor function depends on the fifth coordinate $z$ only,
and $\eta_{\alpha \beta}=\{1,-1,-1,-1\}$ is a Minkowski metric.
Inserting this metric into Eqs.~\eqref{einst} and
\eqref{T}, one 
obtains from the  $\left(^z_z\right)$ component of the Einstein equations
\begin{equation}
\label{eq_gen_p}
p \frac{d^2 p}{d y^2}+\left(\frac{d p}{d y}\right)^2+5 p \frac{d p}{d y}=
\frac{1}{32 p^2 f_{RR}}\left[4p\left(\frac{d p}{d y}+p\right)f_R-\frac{1}{2}f-6p^2\right].
\end{equation}
where the new variable $p=dy/dz$ was introduced,
and the index $R$ denotes the derivative with respect to
the scalar curvature $R$.
Eq.~\eqref{eq_gen_p} is a third-order differential equation
with respect to the metric function $y$,
and the remaining components of the Einstein equations
are fourth-order ones.
Using the expression for
$f(R)$ from \eqref{fR} one can obtain the equation for $y$ in the form
\begin{equation}
\label{eq_n_y}
y^{\prime\prime\prime}-\frac{1}{n}\frac{y^{\prime\prime 2}}{y^{\prime}}
+\left[5-\frac{\frac{7n}{2}-5}{n(n-1)}\right]y^{\prime}y^{\prime\prime}
-\frac{5}{2}\,\frac{n-\frac{5}{2}}{n(n-1)}y^{\prime 3}
=\frac{12 y^{\prime}}{\alpha 8^n n(n-1)}\left(y^{\prime\prime}+\frac{5}{2}
y^{\prime 2}\right)^{2-n},
\end{equation}
where the prime denotes the derivative with respect to $z$.
We note, that by introducing the scaled variable $\bar z= \bar \alpha z$
with $\bar \alpha = \alpha^{\frac{1}{2(1-n)}}$,
Eq.~(\ref{eq_n_y}) becomes independent of $\alpha$.
Thus it is sufficient to solve the equation for $\alpha=1$.
All other solutions are obtained from this solution by scaling.

One can also see from Eq.~(\ref{eq_n_y}) that 
the first derivative $y^{\prime}$ cannot
take the value zero unless
$y^{\prime\prime}(z)= 0$ as well.
As will be shown below there is only one point
in the phase plane where both $y^{\prime}$  and $y^{\prime\prime}$
are equal to zero - the fixed point.



Due to its complexity, we will seek numerical solutions of
Eq.~\eqref{eq_n_y}.
But before, let us investigate the qualitative behavior
of solutions of Eq.~\eqref{eq_n_y}.
For this purpose, we rewrite it as a system of three first-order
differential equations
\begin{eqnarray}
\label{sys_first_oder}
y^{\prime}&=&p, \nonumber \\
p^{\prime}&=&v, \\
v^{\prime}&=&\frac{1}{n}\frac{v^2}{p}-\left[5-\frac{\frac{7n}{2}-5}{n(n-1)}\right]p\,v
+\frac{5}{2}\,\frac{n-\frac{5}{2}}{n(n-1)}p^3+\frac{12 p}{\alpha 8^n n(n-1)}\left(v+\frac{5}{2}
p^2\right)^{2-n}.\nonumber
\end{eqnarray}
The fixed point of the system is
\begin{equation}
\label{fix}
\mathcal{A}=\{p\rightarrow 0, v\rightarrow 0\}.
\end{equation}
In order to analyze the behavior of solutions at the fixed point,
let us consider a variation $\delta \ll y$
in the neighborhood of the fixed point for Eq.~\eqref{eq_n_y}.
The corresponding equation for $\delta$
then becomes
\begin{equation}
\label{pert}
\delta^{\prime\prime\prime}-\frac{2}{n}\frac{y^{\prime\prime}}{y^{\prime}}\delta^{\prime \prime}+\frac{1}{n}\frac{y^{\prime\prime 3}}{y^{\prime 2}}\delta^{\prime}=0.
\end{equation}
Taking into account that at the fixed point
denoted by $z=z_{fp}$ the derivatives vanish,
i.e., $y^{\prime}_{fp}=0$ and $y^{\prime\prime}_{fp}=0$,
we seek a solution of Eq.~\eqref{eq_n_y}
in the neighborhood of the fixed point of the form
$$
y=y_{fp}+y^{\prime\prime\prime}_{fp}\frac{(z-z_{fp})^3}{6}.
$$
Then Eq.~\eqref{pert} takes the form
$$
\delta^{\prime\prime\prime}-\frac{4}{n(z-z_{fp})}\,\delta^{\prime \prime}+\frac{4 y^{\prime\prime\prime}_{fp}}{n(z-z_{fp})}\,\delta^{\prime}=0
$$
with the solution
$$
\delta=\delta_0 (z-z_{fp})^{2+4/n}.
$$
To be sure that this solution decays faster than $y$, it is necessary to suppose that
\begin{equation}
\label{n_range}
2+\frac{4}{n}>3 \qquad \Rightarrow \qquad 0<n<4, \quad n\neq 1.
\end{equation}

Let us next analyze the behavior of solutions in the plane
$\{v,p\}$ near the fixed point following the approach of \cite{Reis}.
As an example, let us consider the important case with
$n=2$ (and $\alpha=1$).
In this case Eq.~\eqref{eq_n_y} takes the form
$$
p^{\prime\prime}+f(p,p^{\prime})p^{\prime}+g(p)=0,
$$
or, equivalently,
$$
p^{\prime}=v, \quad v^{\prime}=-g(p)-f(p,v) v,
$$
where
$$
f(p,p^{\prime})=-\frac{1}{2}\frac{p^{\prime}}{p}+4 p, \quad g(p)=\frac{5}{8}p^3-\frac{3}{32}p.
$$
It is obvious from the above expression that
$$
\int\limits_0^{\pm \infty}g(p)dp=G(\pm \infty)=+\infty.
$$
This is a necessary condition to be satisfied in the analysis
of such systems \cite{Reis}.
According to \cite{Reis}, in this case the origin of the coordinates
is then a single fixed point.
It is surrounded by the so-called ``curves of energy''
with the equation
$$
w(p,v)\equiv G(p)+\frac{1}{2}v^2=C.
$$
All of these are closed curves.
In our case the function $g(p)$ is odd,
and therefore these curves are symmetrical
with respect to both coordinate axes.
\begin{figure}[ht]
\begin{minipage}[t]{.49\linewidth}
  \begin{center}
  \includegraphics[height=4.5cm]{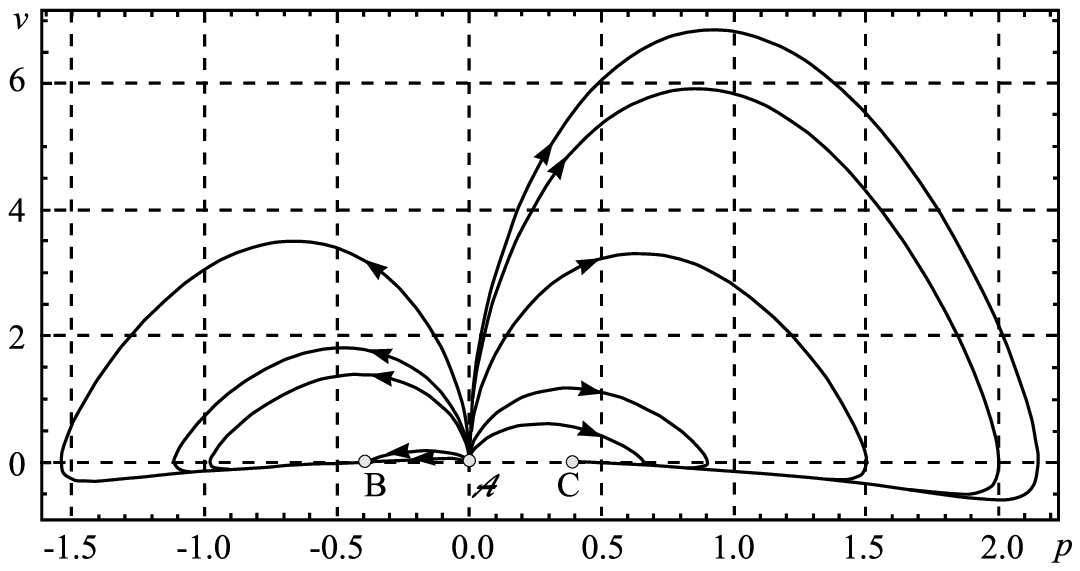}
  \end{center}
\end{minipage}\hfill
\begin{minipage}[t]{.49\linewidth}
  \begin{center}
  \includegraphics[height=4.5cm]{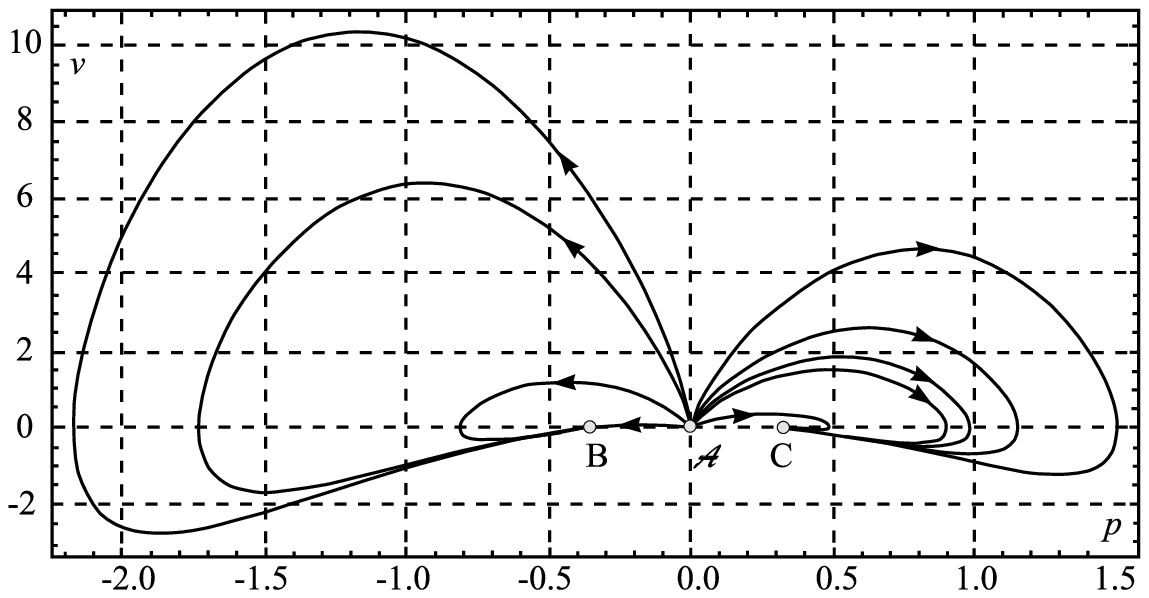}
  \end{center}
\end{minipage}\hfill
\caption{The phase portrait for the case $n=2$ (left panel)
and $n=4/3$ (right panel),
$\alpha=1$. $\mathcal{A}$ is a repulsive node,
$B, C$ denote the asymptotic points \eqref{asymp}
at $z\rightarrow \mp \infty$, respectively.
The different curves correspond to different initial values
of the variable $y$.}
    \label{phaza}
\end{figure}

The behavior of the solutions in the neighborhood of the fixed point
(and the curves of energy) can be estimated as follows:
from the second and third equations of the system \eqref{sys_first_oder},
it is possible to obtain an equation near the fixed point
$$
\frac{dv}{dp}=\frac{1}{2}\frac{v}{p} \quad \Rightarrow \quad v= D\sqrt{p},
$$
where $D$ is an integration constant.
This is a set of curves escaping from the node $p=0, v=0$.
Specifying the boundary conditions in neighborhood
of the fixed point $\mathcal{A}$,
one can then draw a phase portrait of the system
\eqref{sys_first_oder},
presented in Fig.~\ref{phaza}.
As a further example, the model with $n=4/3$ is also presented in this figure.
\begin{figure}[ht]
\centering
  \includegraphics[height=5cm]{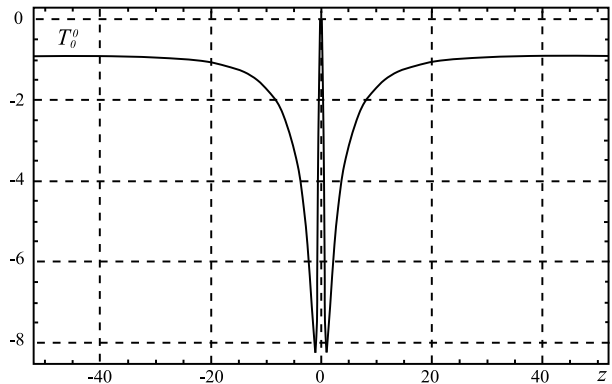}
  \caption{The effective energy density $\hat{T}_{0}^0$
for the case $n=2$, $\alpha=1$.
The point $\hat{T}_{0}^0=0$ has been placed at $z=0$
by shifting $z\rightarrow z-z_0$.}
    \label{energy_n_2}
\end{figure}

The asymptotical form of the solution for arbitrary $n$ is given by
\begin{equation}
\label{asymp}
y_{\infty}=  k_n |z|, \quad k_n=\left[\frac{12}{\alpha (1-\frac{2 n}{5}) 20^n}\right]^{\frac{1}{2(n-1)}}.
\end{equation}
One can see that there is an upper bound for the parameter $n$,
$n<5/2$.
Taking into account the condition from \eqref{n_range}
one can conclude that such a type of thick brane solution
can exist only in the range
$0<n<5/2, n\neq 1$.

Let us now discuss the corresponding distribution
of the effective energy density
$\hat{T}_{0}^0=-3\left(y^{\prime\prime}+2 y^{\prime 2}\right)$
for the case $n=2$,
shown in Fig.~\ref{energy_n_2}.
Asymptotically $\hat{T}_{0}^0$ goes to a constant negative value:
$\hat{T}_{0 (\pm\infty)}^0\rightarrow -6 y^{\prime 2}_{\infty}=-0.9$,
and the 5-dimensional scalar curvature
$R^{(5)}=8 y^{\prime \prime}+20 y^{\prime 2}$
goes to a positive constant, $R^{(5)}_{(\pm\infty)}=3$.
Thus this corresponds to an asymptotical anti-de Sitter solution.


\section{Trapping of matter}
In this section we consider the trapping of a test scalar field
on the brane considered above.
For this purpose, let us use the approach
suggested in the paper \cite{Abdyrakhmanov:2005fs}.
We consider the test complex scalar field $\chi$ with the Lagrangian
$$
L_{\chi}=\frac{1}{2}\partial_A \chi^{*}\partial^A\chi-\frac{1}{2}m_0^2 \chi^*\chi,
$$
where $m_0$ is the mass of the test field.
Using this Lagrangian, we find the equation for the scalar field
\begin{equation}
\label{tf_eq}
\frac{1}{\sqrt{-^5 g}}\frac{\partial}{\partial x^A}\left(\sqrt{-^5 g} g^{AB}\frac{\partial \chi}{\partial x^B}\right)=-m_0^2 \chi.
\end{equation}
Here $\chi$ is a function of all coordinates, $\chi=\chi(x^A)$.
Taking into account that the canonically conjugate momenta
$p_\mu=(E, \vec{p}\,)$ are integrals of motion,
we will seek a solution in the form
$$
  \chi (x^{A}) = X(z) \exp (-ip_{\mu }x^{\mu }).
$$
Inserting this ansatz into Eq.~\eqref{tf_eq},
leads to the equation for $X(z)$
$$
    X^{\prime\prime} + \sqrt{-^5g} (p^{\mu }p_{\mu } -m_{0}^2 ) X = 0,
$$
or, taking into account that
$p^{\mu}p_{\mu}=e^{-2 y}\left(E^2-\vec{p}\,^2\right)$,
we obtain
$$
X^{\prime\prime} + \left[\left(E^2-\vec{p}\,^2\right)e^{2 y}
 -m_0^2 e^{4 y}\right] X = 0.
$$
According to Eq.~\eqref{asymp},
asymptotically $y_{\infty}= k_n |z|, \,\,k_n>0$.
That is why the dominant term in the above equation
will be the term with $e^{4 y}$.
Thus
$$
X^{\prime\prime}-m_0^2 e^{4 k_n |z|}  X = 0
$$
with the asymptotically decaying solution
\begin{equation}
\label{asymp_sol}
X_{\infty}\approx C \sqrt{\frac{2 k_n}{m_0}} e^{-2 k_n |z|} \exp \left(-\frac{m_0}{2k_n}e^{2k_n |z|}\right),
\end{equation}
where $C$ is an integration constant.

As a necessary condition for the trapping of matter on the brane,
one should require finiteness of the field energy
per unit 3-volume of the brane, i.e.,
\begin{equation}
\label{E_tot}
     E_{\rm tot}[\chi] = \int \limits_{-\infty}^{\infty} T^0_0 \sqrt{-^5g} \,dz
          = \int\limits_{-\infty}^{\infty}
            e^{4 k_n |z|}\left[ e^{-2 k_n |z|}(E^2 + \vec{p}\,^2)X^2
	    + m_0^2 X^2 + X^{\prime 2} \right] dz < \infty,
\end{equation}
and also finiteness of the norm of the field $\chi$
$$
||\chi||^2 = \int\limits_{-\infty }^{\infty }\sqrt{-^5g}\,\chi^*\chi\,dz
    	       = \int_{-\infty }^{\infty } e^{4 k_n |z|}\,X^2 \,dz.
$$
From the solution \eqref{asymp_sol} it is evident
that both $E_{\rm tot}$ and $||\chi||$ converge asymptotically.

\section{Conclusion}

We have considered the 5-dimensional thick brane model
in $f(R)\sim R^n$ theory.
It was shown that
regular solutions with asymptotic \eqref{asymp} exist
in the range $0<n<5/2,\, n\neq 1$.
Space is asymptotically anti-de Sitter in such a brane model.

A special feature of the model is the presence of the fixed point
$\mathcal{A}$.
Since Eq.~\eqref{eq_n_y} is invariant under the
shift of the independent variable $z\rightarrow z+z_0$,
the position of the brane is arbitrary
and it can be placed at any point
on the axis $z$, including $z=0$.
The most natural assumption seems to be
that the brane is situated at the fixed point
$\mathcal{A}$ at $z=0$.
Then 
$y(0)=const, \, y^{\prime}(0)=0,\, y^{\prime\prime}(0)=0$
holds on the brane.

The presence of the repulsive fixed point $\mathcal{A}$ (node)
in the system allows to set the boundary conditions arbitrarily
in the neighborhood of the fixed point.
Anyway, all solutions will escape from $\mathcal{A}$
and tend to the asymptotic value \eqref{asymp}.
It allows to not provide any special conditions
for the model parameters (fine-tuning conditions)
that is typically necessary for other brane models
(see, for example, Ref.~\cite{Abdyrakhmanov:2005fs}).

Consideration of the behavior of a test scalar field
in the bulk has shown that such a field is trapped by the $f(R)$-brane.
Note that the trapping is purely gravitational.

For $n=2$, this $f(R)$ theory has also been employed
for the thin brane model \cite{Parry:2005eb}.
In the vicinity of the brane
(located at $z=0$) a similar behaviour
of the metric function is found, $y(z) \sim z^3$.
However, at some finite value $z_s$, 
a singularity is encountered.
Thus the thin brane model appears to be plagued by bulk
singularities, very much in contrast to the
thick brane model considered in this paper.

\section*{Acknowledgements}
V.D. is grateful to the Research Group Linkage Programme of the Alexander von
Humboldt Foundation for the support of this research. V.F. was funded by 
the research grant from the German Academic Exchange Service  (DAAD).

\end{document}